\documentclass[aps,pra,reprint,floatfix,superscriptaddress,showpacs,longbibliography]{revtex4-1}
\usepackage{amsmath,amsthm,amsfonts,amssymb}
\usepackage{bm}
\usepackage[colorlinks=true,citecolor=blue,linkcolor=blue,urlcolor=blue]{hyperref}
\usepackage[usenames,dvipsnames]{xcolor}
\usepackage{longtable}
\usepackage{graphicx}
\usepackage{float}
\usepackage{placeins}

\def\ketm#1{  \left\vert  #1   \right\rangle   }

\def\sprm#1#2{  \left\langle #1 \left\vert \right. #2 \right\rangle   }

\def\rmem#1#2#3{  \left\langle #1 \left\vert \left\vert  #2
                  \right\vert \right\vert #3 \right\rangle   }

\def\sixjm#1#2#3#4#5#6{  \left\{ \begin{array}{ccc}
                                               #1 & #2 & #3  \\
                                               #4 & #5 & #6
                     \end{array} \right\}   }
\begin{document}

\title{Polarization measurement of dielectronic recombination transitions in highly charged krypton ions}

\author{Chintan~Shah}\email{chintan@mpi-hd.mpg.de}
\affiliation{Physikalisches Institut der Universit\"at Heidelberg, 69120 Heidelberg, Germany}

\author{Holger~J\"org}
\affiliation{Physikalisches Institut der Universit\"at Heidelberg, 69120 Heidelberg, Germany}

\author{Sven~Bernitt}
\affiliation{Max-Planck-Institut f\"ur Kernphysik, 69117 Heidelberg, Germany}
\affiliation{Institut für Optik und Quantenelektronik, Friedrich-Schiller-Universität, 07743 Jena, Germany}

\author{Stepan~Dobrodey}
\affiliation{Max-Planck-Institut f\"ur Kernphysik, 69117 Heidelberg, Germany}

\author{Ren\'e Steinbr\"uge}
\affiliation{Max-Planck-Institut f\"ur Kernphysik, 69117 Heidelberg, Germany}

\author{Christian~Beilmann}
\altaffiliation{Now at Karlsruhe Institute of Technology, 76131 Karlsruhe, Germany}
\affiliation{Physikalisches Institut der Universit\"at Heidelberg, 69120 Heidelberg, Germany}
\affiliation{Max-Planck-Institut f\"ur Kernphysik, 69117 Heidelberg, Germany}

\author{Pedro~Amaro}
\altaffiliation{Now at LIBPhys-UNL,~Departamento de F\'isica,~FCT-UNL,  P-2829-516, Caparica, Portugal}
\affiliation{Physikalisches Institut der Universit\"at Heidelberg, 69120 Heidelberg, Germany}

\author{Zhimin~Hu}
\affiliation{Physikalisches Institut der Universit\"at Heidelberg, 69120 Heidelberg, Germany}

\author{Sebastian~Weber}
\affiliation{Physikalisches Institut der Universit\"at Heidelberg, 69120 Heidelberg, Germany}

\author{Stephan~Fritzsche}
\affiliation{Helmholtz-Institut Jena, 07743 Jena, Germany}
\affiliation{Theoretisch-Physikalisches Institut, Friedrich-Schiller-Universit\"at Jena, 07743 Jena, Germany}

\author{Andrey~Surzhykov}
\affiliation{Helmholtz-Institut Jena, 07743 Jena, Germany}

\author{Jos\'e~R.~{Crespo L\'opez-Urrutia}}
\affiliation{Max-Planck-Institut f\"ur Kernphysik, 69117 Heidelberg, Germany}

\author{Stanislav~Tashenov}
\affiliation{Physikalisches Institut der Universit\"at Heidelberg, 69120 Heidelberg, Germany}

\begin{abstract}
We report linear polarization measurements of x rays emitted due to dielectronic recombination into highly charged krypton ions. The ions in the He-like through O-like charge states were populated in an electron beam ion trap with the electron beam energy adjusted to recombination resonances in order to produce \emph{K}$\alpha$ x rays. The x rays were detected with a newly developed Compton polarimeter using a beryllium scattering target and 12 silicon x-ray detector diodes sampling the azimuthal distribution of the scattered x rays. The extracted degrees of linear polarization of several dielectronic recombination transitions agree with results of relativistic distorted--wave calculations. We also demonstrate a high sensitivity of the polarization to the Breit interaction, which is remarkable for a medium-\emph{Z} element like krypton. The experimental results can be used for polarization diagnostics of hot astrophysical and laboratory fusion plasmas.
\end{abstract}

\date{7 October 2015}


\maketitle

%
\section{Introduction}\label{sec:intro}
%
Collisions of energetic electrons with highly charged ions (HCI), abundant in hot plasmas, may lead to emission of anisotropic and polarized characteristic x rays due to an anisotropy of the electron momentum distribution. 
Measurement of polarization of x-ray lines can provide information on the directionality of the electron currents and the orientation of the magnetic field lines in both hot astrophysical and laboratory plasmas~\cite{haug1979, inal1987}. Such diagnostics requires detailed knowledge of polarization properties of atomic processes leading to the x-ray emission.



Dielectronic recombination (DR)~\cite{massey1942} is a resonant process in which a free electron is captured into an ion under the simultaneous excitation of a bound electron and the intermediate excited state is stabilized by emitting an x ray. DR is one of the dominant recombination mechanisms in hot plasmas: its cross section at the resonance energy is often orders of magnitude larger than that of competing recombination processes~\cite{zhang2004}. It strongly affects both the charge state distribution and the x-ray spectrum~\cite{burgess1964, lagattuta1997, gu2001}. Therefore, an accurate knowledge of DR cross sections, and likewise of the polarization of the DR x rays is needed for modeling of astrophysical and fusion plasmas~\cite{dubau1980, jacobs1997}.

Apart from applications in plasma diagnostics, DR was found to be sensitive to fine details of the electron--electron interaction. Namely, magnetic interactions and a retardation in the exchange of virtual photons between the electrons, commonly referred as the Breit interaction~\cite{breit1929}, affect the resonance strengths of DR transitions~\cite{nakamura2008}, as well as the angular distribution and polarization of the emitted x rays~\cite{chen1995, gail1998, fritzsche2009, fritzsche2011}.

\begin{figure*}
    \includegraphics[clip=true,width=0.8\textwidth]{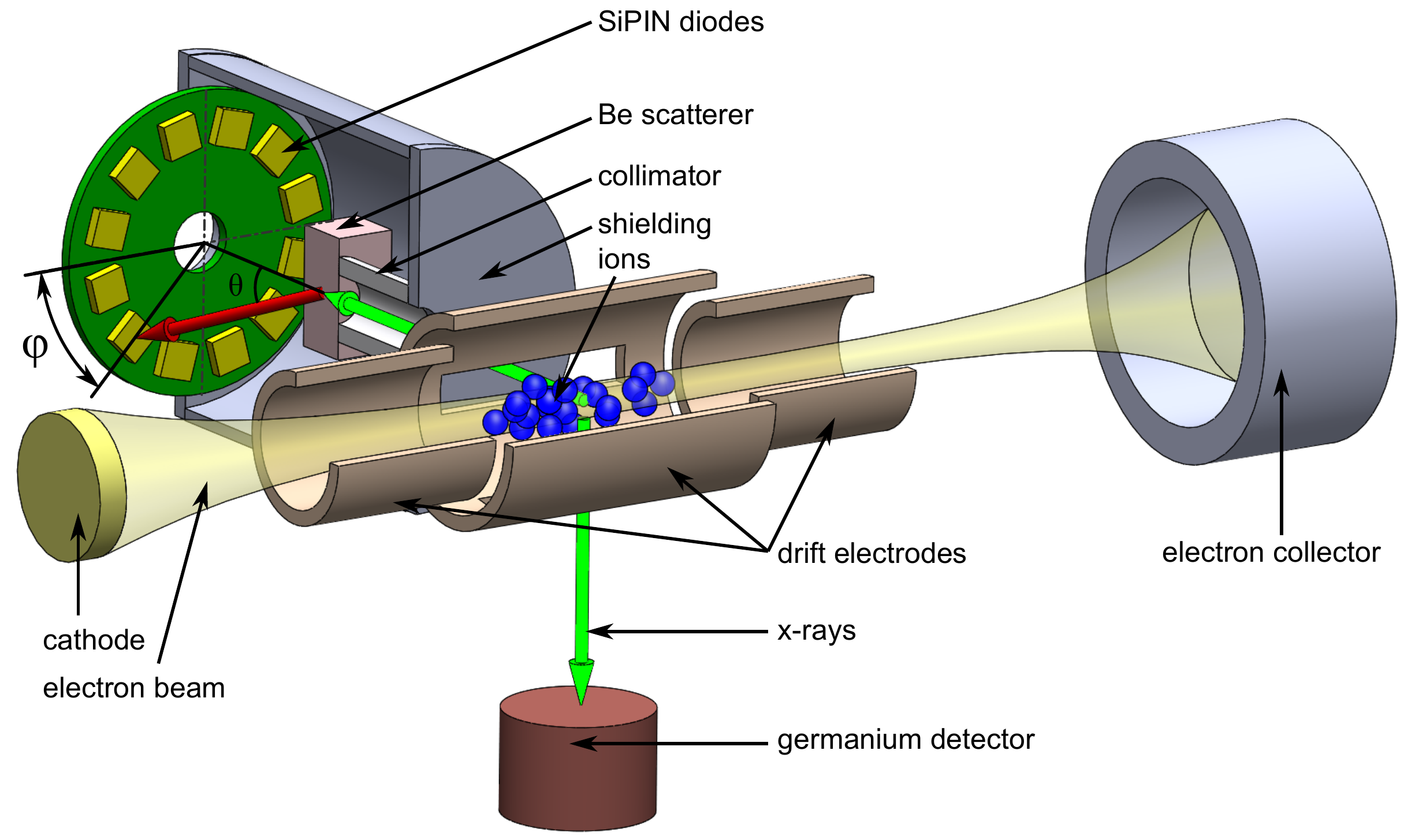}
    \caption{(Color online) Schematic diagram of the EBIT and the layout of the Compton polarimeter. The electrons are emitted from the cathode and accelerated towards the trap center where they collide with trapped highly charged krypton ions. The x rays, emitted in the collision processes, are collimated and Compton-scattered in a block of beryllium and their azimuthal scattering distribution is sampled by an array of SiPIN diodes. An additional germanium detector is used to observe unscattered x rays.} 
    \label{fig:setup}
\end{figure*}

Experimental studies of DR resonances with collision energies of several keV can now be carried out using electron beam ion traps (EBIT). Such devices can produce HCIs in a broad range of charge states. Their interactions with a mono-energetic and unidirectional electron beam were in general studied by observing x rays emitted perpendicular to the electron beam axis. The interpretation of those experimental results require accurate knowledge of the x-ray angular distribution and polarization~\cite{fuchs1998, radtke2000, yao2010, hu2013, xiong2013}. Recently, the angular distribution of DR x rays was measured for one electronic transition in Li-like praseodymium, holmium and gold ions~\cite{hu2012, hu2014}. Polarization properties of dielectronic satellite lines of Li-, Be- and B-like iron ions were also studied~\cite{shlyaptseva1997, shlyaptseva1998, shlyaptseva1999}.
Furthermore, polarization measurements of x-ray lines produced by electron impact excitation were performed for a wide range of elements~\cite{beiersdorfer1996, takacs1996, beiersdorfer1997, beiersdorfer1999, nakamura2001, robbins2004, robbins2006}. In those experiments, the polarization sensitivity of a Bragg crystal spectrometer was used to obtain the polarization-dependent spectra. Although these spectrometers have excellent energy resolutions, their efficiencies are typically small. 
Another x-ray polarimetry technique uses Compton scattering and position sensitive solid-state detectors. This technique has a high efficiency and it is applicable to a wide range of x-ray energies. Furthermore, Compton polarimeters can measure both the degree and the angle of polarization. They were used in studying the bremsstrahlung~\cite{tashenov2011, martin2012, tashenov2013} and radiative electron capture~\cite{tashenov2006, weber2010}. 

Recently, Compton polarimetry was applied at an EBIT for studying the DR into highly-charged xenon ions~\cite{joerg2015}. In this paper, we extend these investigations to a medium-\emph{Z} element krypton and present a detailed data analysis and theoretical calculations of the degree of linear polarization of DR x rays. We have chosen krypton because it was proposed as a promising candidate for x-ray diagnostic measurements in order to determine the central ion temperature in the ITER plasma~\cite{bitter1993, widmann1995}.
The highly charged krypton ions in the He-like through O-like charge states were produced in an EBIT. In this trap, the electron beam energy was tuned to the centroids of the well-resolved \emph{K}-shell DR resonances. For monitoring the x-ray spectrum and adjusting the electron beam energy, the x rays were registered by a germanium detector. The linear polarization of x rays emitted during radiative stabilization of an excited state was measured by a recently developed Compton polarimeter, which consists of a scattering target and an array of silicon PIN (SiPIN) diodes~\cite{Helmuthbook} to sample the azimuthal distribution of the scattered x rays. We have measured the polarization of x rays following the formation of five DR resonances as well as due to radiative recombination (RR) of electrons into the $n=2$ shell of highly charged krypton ions. The measured degree of linear polarization of \emph{K}-shell x-ray lines agrees well with the theoretical calculations carried out using the relativistic configuration interaction based Flexible Atomic Code (\textsc{FAC})~\cite{gu2008}. 


%
\section{Experiment}\label{sec:exp}
%

The FLASH-EBIT~\cite{epp2007, epp2010} developed at Max-Planck-Institute for Nuclear Physics was used for the present experiment, see Fig.~\ref{fig:setup}. In this experiment, an electron beam is emitted from a thermionic cathode with an intensity of 100~mA and it is guided by a 6-T magnetic field through a set of drift tubes. According to the Herrmann's theory~\cite{herrmann1958} the beam is compressed by the magnetic field to a diameter of $\sim$~50~$\mu$m. An electrostatic trap in the axial direction is created by biasing the middle drift tube to a lower potential, as compared to the neighboring drift tubes. The space charge potential of the highly compressed electron beam traps the ions in the radial direction. The energy of the electrons passing through the trap region is defined by the potential difference between the cathode and the central drift tube.

Krypton gas was continuously injected into the trap by means of an atomic beam using a two-stage collimation and differential pumping system. Highly charged krypton ions in the He-like through O-like charge states were produced through successive electron impact ionization. The trapped ions are dumped periodically from the trap to reduce the accumulation of unwanted heavy ionic species emitted from the electron gun, such as barium and tungsten.

We swept the electron beam energy across the range of \emph{KLL} DR resonances, where a free electron is captured into the \emph{L}-shell while exciting a bound electron from the \emph{K}- to the \emph{L}-shell of the ion. The electron beam energy varied from 8.7 to 9.8~keV with a slow slew rate of 1.8~eV/s to ensure an equilibrium charge state distribution~\cite{knapp1993}. The experimental parameters were optimized to get a high x-ray count rate while allowing for the electron beam energy resolution sufficient to resolve the DR resonances. For a moderate beam current of 100~mA and an axial trap depth of about 100~V, we achieved a collision energy resolution of 21-eV full width at half maximum (FWHM) at 9~keV.

\begin{figure}
    \includegraphics[clip=true,width=0.98\columnwidth]{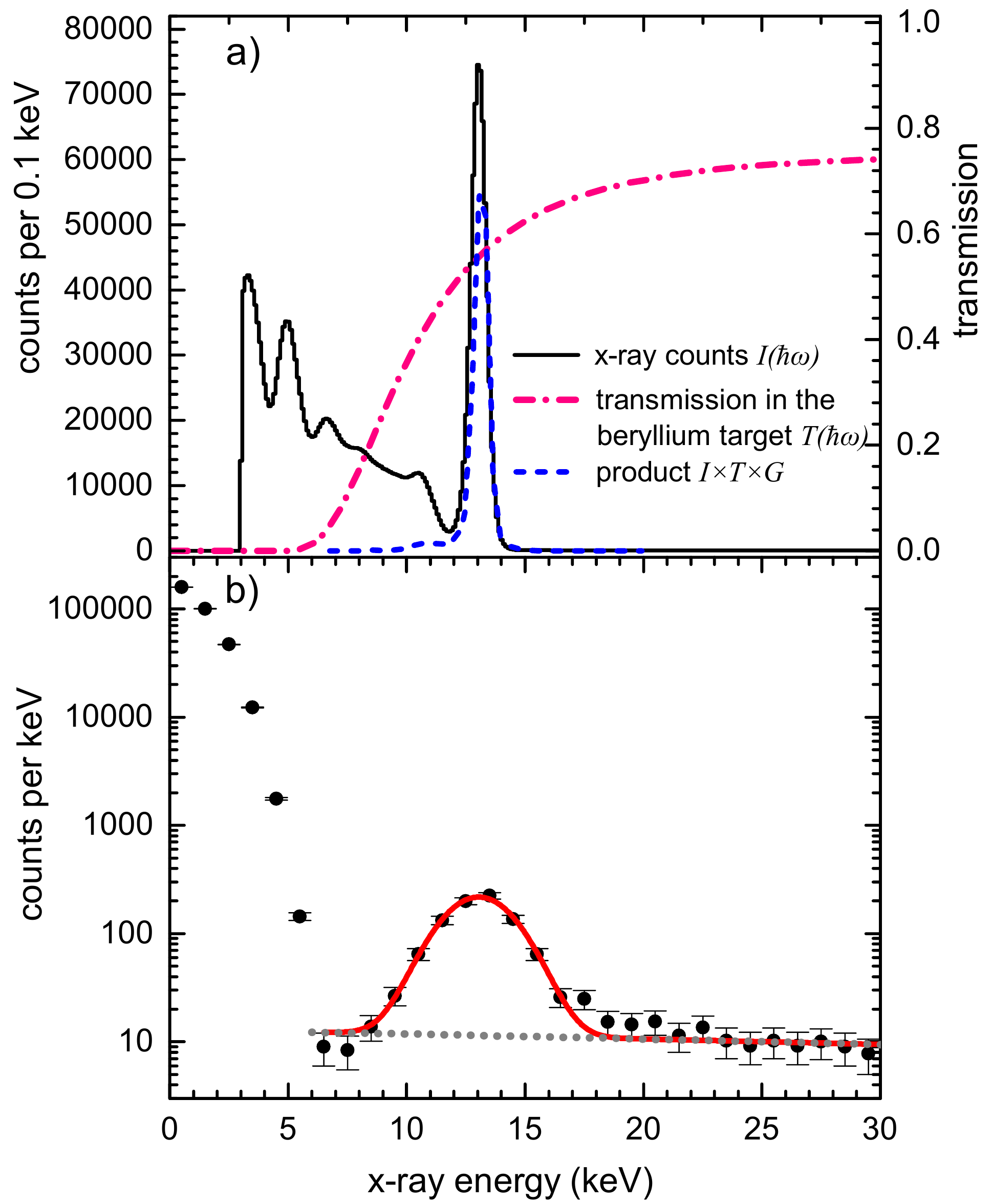}
    \caption{(Color online) (a) An x-ray spectrum~\emph{I($\hbar\omega$)} obtained with the germanium detector (solid line). The dash-dotted line is the energy dependent x-ray transmission probability \emph{T($\hbar\omega$)} through the 10 mm thick beryllium scattering target. The dashed line is the product of the both previous functions and the response function~\emph{G($\hbar\omega$)} of the SiPIN diode to 13~keV x rays: 
$G(\hbar\omega) = \exp\left(-4\ln 2\left(\frac{\hbar\omega - 13~\mathrm{keV}}{3.2~\mathrm{keV}}\right)^2\right)$
. (b) A typical x-ray energy spectrum observed by a SiPIN diode when the electron beam energy is tuned to a DR resonance. The solid curve represents the fitted DR peak having the profile of \emph{G($\hbar\omega$)} and the background is shown as the dashed line.}
    \label{fig:pol_x-ray}
\end{figure}

\begin{figure*}
    \includegraphics[clip=true,width=0.9\textwidth]{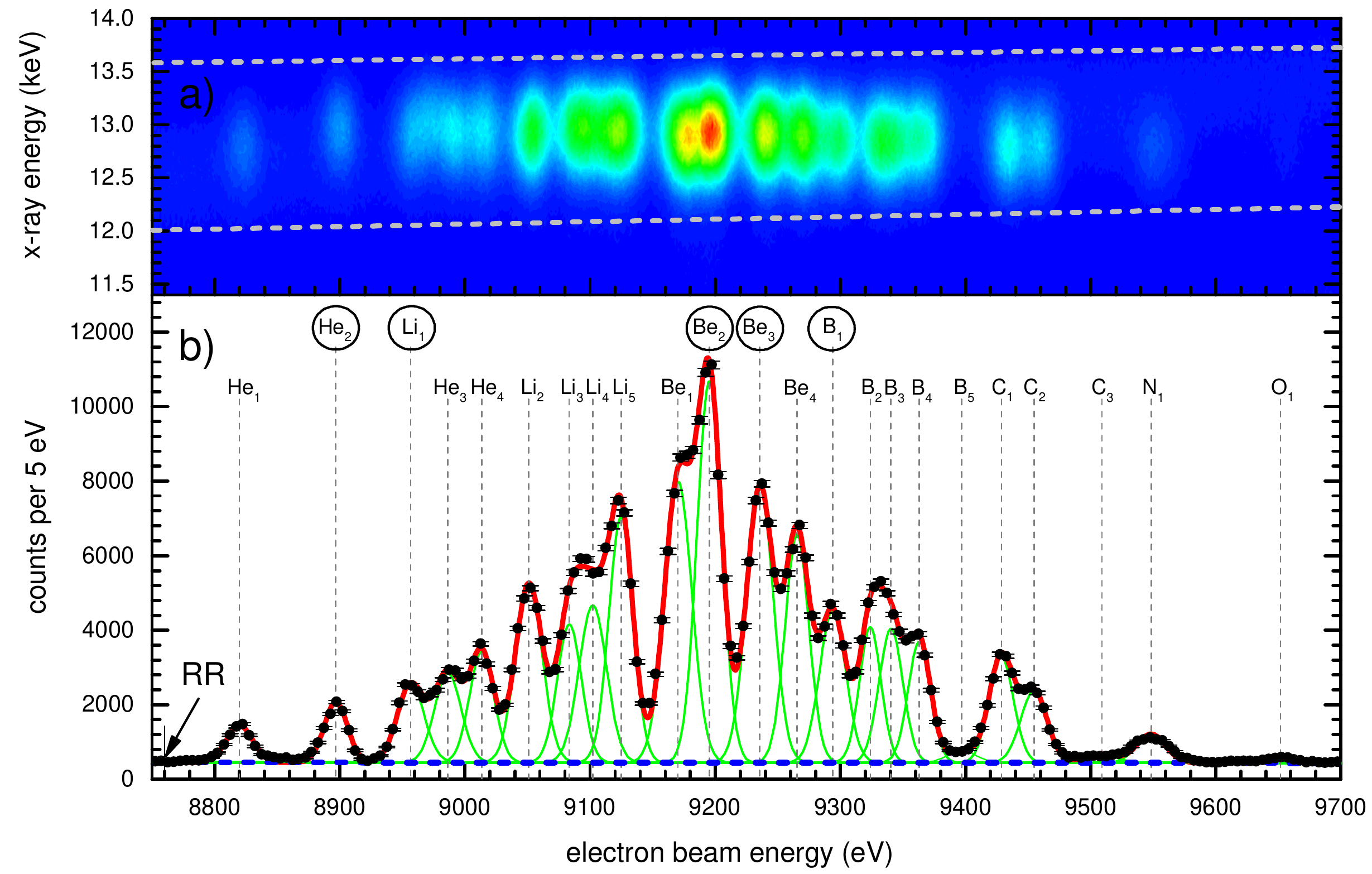}
    \caption{(Color online) (a) Intensity of x rays registered by the germanium detector as a function of the x-ray and the electron beam energies. The dashed lines indicate a region of the \emph{KLL} DR resonances. (b) Intensity of the x rays within the outlined energy window as a function of the electron beam energy. The blue dashed line represents the RR background, the thin green solid lines correspond to the fits of the individual \emph{KLL} DR resonances, and the thick red solid line represents the fit of the complete spectrum. The resonances are labeled by their initial charge state followed by a number. The x-ray polarization was measured for the resonances marked by circles.} 
    \label{fig:spectra}
\end{figure*}

A high-purity germanium detector with an energy resolution of 750-eV FWHM at 13~keV was used to detect the x rays emitted perpendicularly to the electron beam propagation direction. Fig.~\ref{fig:pol_x-ray} (a) shows a typical x-ray energy spectrum observed by the germanium detector. At lower energies, the spectrum is dominated by bremsstrahlung and x-ray lines produced by the trapped impurity ion species. The weak feature at 10~keV is produced by RR into the $n=3$ shell of Kr HCI. At 13~keV, the x rays are produced by RR into the $n=2$ shell and by the \emph{KLL} DR processes. The intensity of x rays as a function of the x-ray and electron beam energies is shown in Fig.~\ref{fig:spectra} (a), where the diagonal window outlines the region of RR into the $n=2$ shell and \emph{KLL} DR resonances. Several bright spots at well-defined electron beam energies correspond to the x-ray intensity enhancements due to DR resonances in different ionic states of krypton.

%
%
\begin{figure*}
    \includegraphics[clip=true,width=0.95\textwidth]{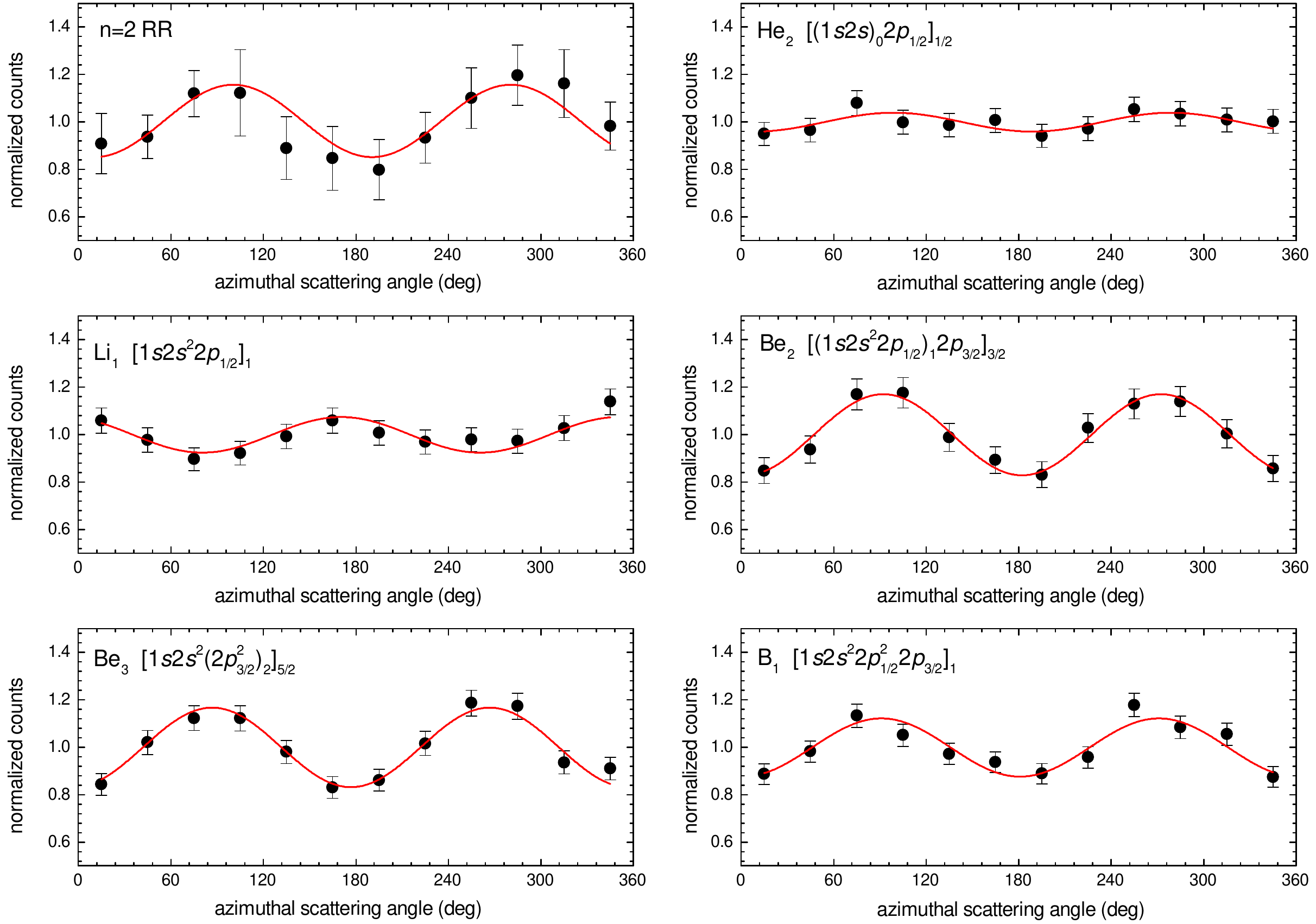}
    \caption{(Color online) Measured azimuthal distributions of the Compton-scattered x rays produced by RR of free electrons into the $n=2$ shell and due to the \emph{KLL} DR resonances in krypton ions. The solid lines represent the fits of Eq.~\eqref{eq:fitfunc} to these distributions.} 
    \label{fig:fitresult}
\end{figure*}
%


The linear polarization of x rays emitted perpendicular to the electron beam axis in RR and DR processes is measured by the dedicated Compton polarimeter optimized for x-ray energies of 10--30~keV, see Fig.~\ref{fig:setup}. The detailed design and the operation principles of this polarimeter will be published elsewhere~\cite{weber_nim2015}. Briefly, the beam of polarized x rays is collimated by a round aperture with a diameter of 15~mm and scattered by 10~mm thick beryllium target. Beryllium is chosen for its low photo-absorption cross section. The energy dependent x-ray transmission of the beryllium scatterer is shown in Fig.~\ref{fig:pol_x-ray} (a). Collimated x rays are scattered at a polar angle of $\theta$ $\approx$ 50$^\circ$ and registered by 12 SiPIN diodes with dimensions of 7~$\times$~7~$\times$~0.38~mm, providing the azimuthal scattering intensity distribution. The SiPIN diodes are operated at room temperature, their signals are amplified with charge sensitive preamplifiers and read out using 100~MHz sampling 14-bit analog-to-digital converters. The moving window deconvolution algorithm~\cite{georgiev1994} is used to extract x-ray energies. This setup provided the energy resolution of 3.2-keV FWHM at 13~keV. Fig.~\ref{fig:pol_x-ray} (b) shows a typical x-ray spectrum observed by a SiPIN diode when the electron beam energy is tuned to the DR resonance. The peak at 13~keV corresponds to the x rays emitted by RR into the $n=2$ shell and \emph{KLL}~DR processes. Since SiPIN diodes have a lower energy resolution compared to the germanium detector, the x rays due to RR into the $n=3$ shell and some bremsstrahlung x rays also contribute to the observed intensity. The feature below 6~keV is due to electronic noise. 

%
\section{Data analysis}\label{sec:data}
%
The total number of x rays detected within the outlined window in the upper panel of Fig.~\ref{fig:spectra} is shown in the lower panel of the same figure as a function of the electron beam energy. 
The observed \emph{KLL}~DR resonances are fitted with a set of Gaussian functions. In the fitting procedure, the centroids and the amplitudes of the resonances are treated as free parameters while their widths are fixed to the experimental electron beam energy resolution of 21~eV. 
The resonances were identified with the help of theoretical calculations performed with \textsc{FAC}. They are then labeled according to the initial charge state of the ion (before the capture of the electron) followed by a number. 
The electron beam energy was calibrated using He$_1$ ($[1s2s^2]_{1/2}$) and C$_1$ ($[1s2s^2 (2p_{1/2} )^2 (2p_{3/2} )^2 ]_{5/2}$) resonances, identified
by the intermediate excited states, having theoretical resonance energies of 8820.6~eV and 9430.5~eV, respectively. An overall good agreement is obtained between the measured and the calculated resonance energies.
We measured the polarization of x rays produced by the well-resolved He$_2$, Li$_1$, Be$_2$, Be$_3$ and B$_1$ DR resonances as well as by the RR process. The x rays produced by the He$_1$ resonance are expected to be unpolarized; we used this resonance to control possible systematic effects.


For the incoming x rays with the energy $\hbar\omega$ the Compton scattering differential cross section is given by the Klein-Nishina formula~\cite{tashenov2006},
\begin{equation} 
\frac{d\sigma}{d\Omega} \propto \frac{\hbar\omega}{\hbar\omega^\prime}+\frac{\hbar\omega^\prime}{\hbar\omega} - \sin^2\theta - P \sin^2\theta \, \cos2(\varphi-\varphi_{0}),
\label{eq:klein-nishina}
\end{equation}
where $\hbar\omega^\prime$ is the energy of the scattered x rays, $P$ and $\varphi_0$ are the degree and the angle of linear polarization and $\theta$ and $\varphi$ are the polar and the azimuthal scattering angles. Here $\varphi$ and $\varphi_0$ are measured relative to the reaction plane defined by the electron beam and the x-ray propagation directions. Since the geometry of the radiative collision process is symmetric with respect to the reaction plane, the angle of linear polarization $\varphi_{0}$ vanishes. In this case the degree of linear polarization $P$ is identical to the first Stokes parameter $P = (I_{\parallel}-I_{\perp})/(I_{\parallel}+I_{\perp})$, where $I_{\parallel}$ and $I_{\perp}$ are the intensities of x rays polarized within the reaction plane and perpendicular to it.

The intensities of the scattered DR x rays observed by different SiPIN diodes are extracted by fitting the corresponding x-ray spectra with Gaussian functions together with the background approximated by a linear function, see Fig.~\ref{fig:pol_x-ray} (b). The line intensity, width and centroid as well as the offset and the slope of the background were treated as free parameters. The intensities of the scattered DR x rays observed by each individual SiPIN diode are shown in Fig.~\ref{fig:fitresult} forming a function of the azimuthal scattering angle $\varphi$. The intensity modulation indicates the linear polarization of the incoming x rays. According to Eq.~\eqref{eq:klein-nishina}, it was fitted with the function:
\begin{equation}
I(\varphi) \propto  1 - P \, M \cos 2(\varphi - \varphi_0).
\label{eq:fitfunc}
\end{equation}
Although the angle $\varphi_0$ should vanish, here it takes a possible misalignment of the EBIT and the polarimeter axes into account. The factor $M$ represents the modulation produced by a beam of 100~\% polarized x rays. 
It was obtained by Monte-Carlo simulations using the \textsc{GEANT}4 toolkit~\cite{agostinelli2003, allison2006} together with the Livermore low-energy electromagnetic model for the scattering processes~\cite{depaola2003}. The simulations take Compton scattering, Rayleigh scattering, photoelectric effect and bremsstrahlung into account along with the dimensions of the collimator, the scatterer and the SiPIN diodes~\cite{weber_nim2015}. They confirm that the azimuthal scattering distribution of 13~keV x rays has a form of Eq.~\eqref{eq:fitfunc} with $M$ = 0.374. 






Due to the low energy resolution of the SiPIN diodes, the degree of polarization, extracted from the azimuthal scattering distribution, arises from contributions from \emph{KLL}~DR, RR and bremsstrahlung. 
To quantify these effects, first we have estimated the contributions of bremsstrahlung and RR into $n \geq 3$ shells to the intensity of the x-ray line observed by the SiPIN diodes. For this we have used the x-ray spectrum~\emph{I($\hbar\omega$)} recorded by the germanium detector where these contributions are clearly separated from the main x-ray line. By multiplying the germanium x-ray spectrum to the beryllium scatterer transmission probability~\emph{T($\hbar\omega$)} we have obtained the spectrum of the x rays impinging on the SiPIN diodes. A further multiplication by the response function~\emph{G($\hbar\omega$)} of the SiPIN diode took the energy resolution of the diode into account, see the dashed line in Fig.~\ref{fig:pol_x-ray} (a). The contributions of bremsstrahlung and RR into $n \geq 3$ shells to the main x-ray line are visible as a faint feature at lower energies. They modify the x-ray intensity by 1--5~\% and thus they were neglected.

The contribution of RR into $n = 2$ shell to the x-ray intensity measured by the SiPIN diodes cannot be separated from that of \emph{KLL}~DR, even using the x-ray spectrum obtained by the germanium detector. This contribution $f_\mathrm{RR}$, however, was determined by fitting the observed x-ray intensity as a function of the electron beam energy using Gaussian functions for the DR transitions and a linear baseline for RR, see Fig.~\ref{fig:spectra}. The contributions of RR to the DR resonances He$_2$, Li$_1$, Be$_2$, Be$_3$, and B$_1$ were found to be respectively 22 \%, 18 \%, 4 \%, 6 \% and 10 \%, see Table~\ref{tab:result}.

From Eq.~\eqref{eq:klein-nishina} and~\eqref{eq:fitfunc} it follows that the measured degree of polarization $P$ is formed by the polarization of the RR x rays ($P_\mathrm{RR}$) and DR x rays ($P_\mathrm{DR}$):
\begin{equation}
P = P_\mathrm{RR} \, f_\mathrm{RR} + P_\mathrm{DR} \, (1 - f_\mathrm{RR}).
\label{eq:p_DR}
\end{equation}
We measured the RR polarization at a fixed electron beam energy of 8760~eV, which is far from known DR resonances, and found $P_\mathrm{RR} = 0.41 \pm 0.1$.
This experimental value was compared with the results of the relativistic distorted--wave calculations performed with \textsc{FAC} for the electrons recombining into $2p_{1/2}$ and $2p_{3/2}$ sub-shells of initially He-like through N-like krypton ions. The theoretical result $P_\mathrm{RR}\approx 0.59$ deviates from the experimental value. We have no explanation for this discrepancy and plan to address it in future experiments.

The degree of polarization of DR x rays $P_\mathrm{DR}$ is extracted by deconvoluting the polarization of RR x rays $P_\mathrm{RR}$ according to Eq.~\eqref{eq:p_DR}. The experimental results are summarized in Table~\ref{tab:result}. 
The uncertainties in the degrees of \emph{KLL} DR polarization $P_\mathrm{DR}$ are caused by the fits of Eq.~\eqref{eq:fitfunc} and the errors of $f_\mathrm{RR}$ and $P_\mathrm{RR}$. Moreover, the depolarization effect introduced by the cyclotron motion of the electrons inside the EBIT was estimated~\cite{gu1999, beiersdorfer2001} and found to be negligibly small compared to the experimental errors.

Using the He$_2$ resonance having the total angular momentum of the intermediate excited state $J_d$ = 1/2 and producing unpolarized x rays, we have verified that our experimental setup does not have instrumental asymmetries affecting the polarization measurement. The modulation of the azimuthal angular distribution of the scattered x rays following He$_2$ resonance seen in Fig.~\ref{fig:fitresult} is due to the admixture of the polarized RR x rays. 

%
%
\section{Theoretical calculations}\label{sec:th} 
%
Since the theoretical analysis of the dielectronic recombination of few--electron ions has been discussed in details previously~\cite{fritzsche2008, fritzsche2009, surzhykov2010, fritzsche2011, hamiltonbook}, here we restrict ourselves to a rather short account of the basic formulas and ideas. In particular, we consider the DR as a two--step process:
\begin{eqnarray}
   \label{eq_DR_description}
   \mathrm{e}^-(E_\mathrm{e} l j) &+& \mathrm{A}^{n+}(\alpha_i J_i) \nonumber \\
   && \hspace*{-1cm} \to \mathrm{A}^{(n-1)+ *}(\alpha_d J_d) \to  \mathrm{A}^{(n-1)+}(\alpha_f J_f) + \gamma \, .
\end{eqnarray}
Here, the first step is the resonant capture of a free electron with energy $E_\mathrm{e}$ and total angular momentum $j$ by an initial ion $\mathrm{A}^{n+}$ in the charge state $n$. This capture leads to the formation of the excited ion $\mathrm{A}^{(n-1)+ *}$, whose charge state is reduced by one and which decays -- in a second step -- under the emission of characteristic x rays. In Eq.~(\ref{eq_DR_description}), moreover, $J_i$, $J_d$ and $J_f$ are the total angular momenta of the initial, intermediate (excited) and final ionic states, and we used $\alpha_i$, $\alpha_d$ and $\alpha_f$ to denote all the additional quantum numbers. 

The resonant capture of an electron by a target ion usually leads to the non--statistical population of excited ionic sublevels $\ketm{\alpha_d J_d M_d}$ with $M_d = -J_d, ... J_d$. Moreover, if the incident electron beam is unidirectional and unpolarized, the residual (excited) ions appear to be aligned which means that their sublevels with the same modulus of the magnetic quantum number $\left|M_d\right|$ are equally populated, while this is not true in general for levels with different moduli. Usually, the alignment is described by the set of parameters $\mathcal{A}_{k}$ that are related to the cross sections $\sigma_{M_d}$ for the capture into the various magnetic sublevels:
\begin{eqnarray}
   \label{eq_alignemnt_definition}
\hspace*{-1.0cm} \mathcal{A}_{k} &=& \sqrt{2J_d + 1} \nonumber \\[0.2cm]
&& \hspace*{-1.0cm} \times \sum\limits_{M_d = -J_d}^{J_d} (-1)^{J_d - M_d} \, \sprm{J_d M_d, \, J_d -M_d}{k 0} \, \frac{\sigma_{M_d}}{\sum\limits_{M^\prime_d} \, \sigma_{M^\prime_d}}\,.
\end{eqnarray}
As seen from this expression and from the properties of the Clebsch--Gordan coefficients $\sprm{....}{..}$, the alignment parameters $\mathcal{A}_{k}$ do not vanish only if $k$ is even and if $k \le 2J_d$. This restriction implies, in particular, that only excited states with $J_d > 1/2$ can be aligned. For example, there is just one non--vanishing parameter $\mathcal{A}_{2}$ for the recombination of a free electron into the intermediate excited state with $J_d$ = 1 or 3/2 and two non--vanishing parameters $\mathcal{A}_{2}$, $\mathcal{A}_{4}$ for $J_d$ = 5/2:
\begin{align}
\mathcal{A}_{2}(\alpha_d J_{d}=1) &= \sqrt{2} \frac{\sigma_{1} -\sigma _{0}}{\sigma _{0} + 2\sigma _{1}},\label{eq:A2_J=1}\\
\mathcal{A}_{2}(\alpha_d J_{d}=3/2) &= \frac{\sigma_{3/2} - \sigma_{1/2}}{\sigma_{1/2} + \sigma_{3/2}},\label{eq:A2_J=3/2}\\
\mathcal{A}_{2}(\alpha_d J_{d}=5/2) &= \sqrt{\frac{1}{14}} \frac{5\sigma_{5/2} - \sigma_{3/2} - 4\sigma_{1/2}}{\sigma_{1/2} + \sigma_{3/2} + \sigma_{5/2}},\label{eq:A2_J=5/2}\\
\mathcal{A}_4(\alpha_d J_{d}=5/2) &= \sqrt{\frac{3}{14}} \frac{\sigma_{5/2} - 3\sigma_{3/2} + 2\sigma_{1/2}}{\sigma_{1/2} + \sigma_{3/2} + \sigma_{5/2}}.\label{eq:A4_J=5/2}
\end{align}
%
%
\begin{table*}[htb]
	\centering
	\caption{The measured degrees of polarization of x-ray transitions following 5 DR resonances. These DR resonances with the centroid energies $E_\mathrm{e}$ are labeled by the initial charge states of the recombining ion followed by a number and identified by the intermediate excited states. The resonances are given in \textit{j--j} coupling notation, where the subscripts after the round brackets stand for the angular momentum of the coupled subshells and those after the square brackets denote the total angular momentum of the state. The factors $f_{RR}$ denote the amount of the RR background admixed to the DR x rays. Experimental uncertainties are given as $1\sigma$. The alignment parameters $\mathcal{A}_2$, the intrinsic anisotropy parameters $\bar\alpha_2^{df}$ and the degrees of x-ray polarization $P_\mathrm{DR}(\mathrm{theory})$ are calculated with \textsc{FAC}. The theoretical results are given for the case of the Coulomb-only electron--electron interaction (C) and for the case of the full inter-electronic interaction with the Breit term included (C+B).}
	\bgroup
	\def\arraystretch{1.4}
	\setlength{\tabcolsep}{0.8em}
	\begin{tabular}{clccrcrrrr}
		\hline \hline
		Label & Excited state & $E_\mathrm{e}$ (eV) & $f_\mathrm{RR}$ & $P_\mathrm{DR}(\mathrm{exp})$ & $\bar \alpha_2^{df}$ & \multicolumn{2}{c}{$\mathcal{A}_2$} & \multicolumn{2}{c}{$P_\mathrm{DR}(\mathrm{theory})$} \\
		&    &    &    &    &    & C  & C+B & C & C+B \\
		\hline
		He$_2$ & $[(1s 2s)_{0} 2p_{1/2}]_{1/2}$ & 8899.5 & 0.22 $\pm$ 0.01 & 0.02 $\pm$ 0.05 & 0.00 & 0.00 & 0.00 & 0.00 & 0.00 \\
		Li$_1$ & $[1s 2s^2 2p_{1/2}]_{1}$ & 8954.0 & 0.18 $\pm$ 0.01 & $-$0.84 $\pm$ 0.05 & 0.68 & 0.70 & 0.64 & $-$0.95 & $-$0.84 \\
		Be$_2$ & $[(1s 2s^2 2p_{1/2})_{1} 2p_{3/2}]_{3/2}$ & 9197.8 & 0.04 $\pm$ 0.002 & 0.45 $\pm$ 0.04 & 0.32 & $-$1.00 & $-$1.00 & 0.42 & 0.42 \\
		Be$_3$ & $[1s 2s^2 (2p_{3/2}^{2})_{2}]_{5/2}$ & 9239.2 & 0.06 $\pm$ 0.003 & 0.46 $\pm$ 0.04 & 0.36 & $-$1.07 & $-$1.07 & 0.48 & 0.48 \\
		B$_1$ & $[1s 2s^2 2p_{1/2}^2 2p_{3/2}]_{1}$ & 9296.8 & 0.10 $\pm$ 0.005 & 0.33 $\pm$ 0.06 & 0.46 & $-$0.69 & $-$0.68 & 0.41 & 0.40 \\
		\hline \hline
	\end{tabular}%
	\egroup
	\label{tab:result}%
\end{table*}%
%

In order to calculate the partial cross sections $\sigma_{M_d}$ and, hence, the alignment parameters (\ref{eq_alignemnt_definition}) we have used the \textsc{FAC} program which treats the resonant capture process within the distorted-wave approximation~\cite{gu2003dr, gu2008}. In this approach, the evaluation of the capture cross sections is traced back to the matrix element $\rmem{\alpha_d J_d}{\hat{V}}{(\alpha_i J_i , \, E_e l j) J}$. This matrix element describes the formation of the excited state $\ketm{\alpha_d J_d}$ due to the electron--electron interaction $\hat{V}$ between the bound electrons in the initial state $\ketm{\alpha_i J_i}$ and the free electron. To better understand the influence of the relativistic Breit interaction on the resonant capture process, we have performed calculations for both the full interaction operator $\hat{V} = \hat{V}^{\rm Coulomb} + \hat{V}^{\rm Breit}$ as well as for just the Coulomb repulsion, $\hat{V} = \hat{V}^{\rm Coulomb}$.

The alignment of excited ionic states $\ketm{\alpha_d J_d}$ can strongly affect the angular distribution and polarization of the characteristic x rays emitted in the second step of the DR process, c.f.~Eq.~(\ref{eq_DR_description}). For example, the linear polarization of the decay x rays is expressed as:
\begin{equation}
   \label{eq_degree_polarization}
P(\theta) = \frac{\sum\limits_{k = 2, 4, ...}  2\,\sqrt{\frac{(k-2)!}{(k+2)!}} \,\, \mathcal{A}_{k} \, g_k \, P^{(2)}_k(\cos\theta)}{1 + \sum\limits_{k = 2, 4, ...} \mathcal{A}_{k} \, f_k \, P_k(\cos\theta)} \, ,
\end{equation}
where the emission angle $\theta$ is defined with respect to the incident electron beam, and where $P_k$ and $P^{(2)}_k$ denote the Legendre and associated Legendre polynomials, correspondingly. Apart from the alignment $\mathcal{A}_{k}$, the polarization $P(\theta)$ depends also on the so--called structure functions $g_k \equiv g_k(\alpha_d J_d, \, \alpha_f J_f)$ and $f_k \equiv f_k(\alpha_d J_d, \, \alpha_f J_f)$ that reflect the electronic structure of the ion. The explicit expressions for these functions are quite elaborate since they contain a summation over the different multipole channels allowed for a particular decay $\ketm{\alpha_d J_d} \to \ketm{\alpha_f J_f} + \gamma$, c.f.~Ref.~\cite{surzhykov2006, surzhykov2010}. For low-- and medium--$Z$ few--electron ions, however, the electric dipole $E1$ channel dominates (if allowed) the radiative decay. In this dipole approximation the structure functions are significantly simplified to:
\begin{eqnarray}
   \label{eq_structure_function}
   f^{(E1)}_{k} &=& -\sqrt{\frac{2}{3}} g^{(E1)}_{k} = \delta_{k, \, 2} \nonumber \\[0.2cm]
   && \hspace{-1cm} \times (-1)^{1 + J_d + J_f} \, \sqrt{\frac{3 (2J_d + 1)}{2}} \, \sixjm{1}{1}{2}{J_d}{J_d}{J_f} \, ,
\end{eqnarray}
and $f^{(E1)}_{k}$ coincides with the anisotropy parameter $\alpha^{df}_2$ introduced earlier in Ref.~\cite{balashovbook}: $f^{(E1)}_{k} = \alpha^{df}_2$. Eq.~(\ref{eq_structure_function}) implies that only the second--rank alignment parameter $\mathcal{A}_{2}$ can affect the polarization of $E1$ photons. Inserting the $f^{(E1)}_{k}$ and $g^{(E1)}_{k}$ into Eq.~(\ref{eq_degree_polarization}) one obtains:
\begin{equation}
   \label{eq_degree_polarization_E1}
   P^{(E1)} = - \frac{3 \mathcal{A}_{2} \, \alpha^{df}_2}{2 - \mathcal{A}_{2} \, \alpha^{df}_2} \, ,
\end{equation}
where we have assumed also that characteristic photons are detected perpendicular to the incident electron beam direction, $\theta = 90^\circ$.

Eq.~(\ref{eq_degree_polarization_E1}) describes the linear polarization of $E1$ photons emitted if the ions undergo a transition between two levels with well-defined total angular momenta and parity. In the present experiment, however, an excited resonance state $\ketm{\alpha_d J_d}$ usually decays into several close--lying different final states $\ketm{\alpha_f J_f}$. Since the energy splitting between these final states is smaller than the energy resolution of our x-ray polarimeter, only the superposition of individual transitions $\ketm{\alpha_d J_d} \to \ketm{\alpha_f J_f} + \gamma$ was observed. For the linear polarization, such a superposition is given by:
\begin{equation}
   \label{p_DR_th}
   P_{\mathrm{DR}} = - \frac{3 \mathcal{A}_{2} \, \bar{\alpha}^{df}_2}{2 - \mathcal{A}_{2} \, \bar{\alpha}^{df}_2} \, ,
\end{equation}
where the ``effective'' anisotropy parameter $\bar{\alpha}^{df}_2 = \sum_f \alpha^{df}_2 A_{r}^{df}/\sum_f A_{r}^{df}$ is obtained upon summation over the unresolved final states and $A_{r}^{df}$ is the rate of the corresponding transition.

%
\section{Results and Discussion}\label{sec:discussion}
%


We have observed that the x rays following the Be$_2$, Be$_3$, and B$_1$ resonances have positive degrees of linear polarization, indicating that they are polarized within the reaction plane. The negative degree of polarization in the case of the Li$_1$ resonance indicates 
that its x rays are polarized perpendicular to the reaction plane. 
A positive degree of polarization can be expected since the angular momentum of the incoming electron in the target frame is perpendicular to the electron collision axis. 
This momentum is transferred into the total angular momentum $J_d$ of the intermediate state predominantly populating magnetic sublevels with the least projection $M_d$.
These are the sublevels with $M_d$ = 0 for integer $J_d$ and $M_d$ = $\pm$ 1/2 for half-integer $J_d$. 
In such cases, according to Eqs.~\eqref{eq:A2_J=1} and \eqref{eq:A2_J=3/2}, the alignment parameter $\mathcal{A}_{2}$ is negative. Hence, for the x-ray transitions with $\bar\alpha_{2}^{df} >$ 0 following the Be$_2$, Be$_3$ and B$_1$ resonances, the degree of linear polarization is positive.
For the contrary case of the Li$_1$ resonance, according to the Ref.~\cite{fritzsche2009}, the population of the magnetic sublevel with $M_d$ = 0 is forbidden in the non-relativistic limit. Thus, the magnetic sublevels with $M_d$ = $\pm$1 are predominantly populated leading to the positive alignment $\mathcal{A}_{2}^d$. Consequently, the degree of linear polarization of x rays following Li$_1$ resonance is negative.

It was shown recently that the Breit interaction between the incident and the target electrons may strongly affect the alignment of the populated states and the polarization of the emitted DR x rays~\cite{chen1995, fritzsche2009, fritzsche2009b}. 
As the Breit interaction occurs as a relativistic correction to the instantaneous Coulomb repulsion, a large contribution is expected especially for high-Z elements. 
In the present work, we here found a quite strong contribution of the Breit interaction to the degree of linear polarization of DR x rays already in krypton, which is significantly lighter than previously studied elements~\cite{hu2012,hu2014}. 
We see an effect of the Breit interaction in the case of Li$_1$ and B$_1$ resonances, while no effect was noticed in the case of the Be$_2$ and Be$_3$ resonances. 
The latter case can be explained by the fact that the electron--electron interaction operator $\hat{V}$ is scalar, and hence it cannot affect the magnetic sublevel population of the excited ion if only one partial wave of the free electron is allowed in the resonant capture process. 
This is the case for the initially Be-like ions $(\mathrm{e}^ {-} + [1s^2 2s^2] _0 \to [(1s 2s^2 2p_ {1/2}) 2p_ {3/2}] _ {3/2, 5/2}) $, where the total angular momentum of the initial state is $J_i$ = 0. Thus, only single partial waves $d_ {3/2}$ and $d_ {5/2}$ are allowed in the formation of Be$_2$ and Be$_3$ resonances, respectively. 
On the other hand, for Li- and B-like initial ions $( \mathrm{e}^{-} + [1s^2 2s]_{1/2} \to [1s 2s^2 2p_{1/2}]_1; \,\, \mathrm{e}^{-} + [1s^2 2s^2 2p_{1/2}]_{1/2} \to [1s 2s^2 2p_{1/2}^2 2p_{3/2}]_1)$, where the total angular momentum of the initial state is $J_i$ = 1/2, two partial waves of the free electron can contribute to the resonant capture process. 
The $p_{1/2}$ and $p_{3/2}$ partial waves are allowed in case of Li$_1$ resonance, while $s_{1/2}$ and $d_{3/2}$ partial waves are allowed in the case of B$_1$ resonance. 
In such instances, the interference between the different allowed partial waves of a free electron governs both the alignment parameters and the polarization of the characteristic radiation. 
This interference depends on the free--bound transition matrix elements and, hence, on the details of the electron--electron interaction~\cite{fritzsche2011, fritzsche2012}. 
The Breit interaction decreases the alignment parameter $\mathcal{A}_2$ of B$_1$ resonance by 1~\% and of Li$_1$ resonance by 9~\%. 
The latter means that in the case of the Li$_1$ resonance the population of the $M_d$=0 sublevel is increased due to the Breit interaction by a factor of 30, which is remarkable for a middle-Z element. 
This noticeable change in the population of the magnetic sublevels leads to a 13 \% decrease in the degree of linear polarization of x rays produced by Li$_1$ resonance. 
Although, this represents a small contribution of the Breit interaction as compared to the previously measured contributions in heavy elements~\cite{hu2014, joerg2015}, this experiment is nonetheless sensitive enough to validate theoretical calculations including the Breit interaction, and to rule out by 2$\sigma$ calculations that treat the electron--electron interaction purely by the Coulomb force.

The external magnetic and electric fields present in the interaction region of an EBIT could modify the DR cross section and magnetic sublevel population~\cite{pindzola1997}. Since in the present experiment the direction of the external magnetic field is the same as that of the electron beam, the magnetic sublevel populations should not be redistributed~\cite{hu2014}. Moreover, small shifts in binding energies due to the Breit interaction do not affect magnetic sublevel populations because the experimental electron--ion collision energy spread is large compared to these shifts.

%
\section{Conclusions}\label{sec:conclu}
%

A good agreement between theoretical predictions and experimental results of polarization of x rays due to \emph{KLL} DR resonances is important for developing reliable plasma polarization diagnostics. Such diagnostics may provide information about the momentum distribution of plasma electrons of specific energies~\cite{inal1989, baranova2003, fujimoto1996}. 
Since several dominant \emph{KLL} resonant recombination transitions in krypton ions were studied in the present work, the results are relevant for fusion plasmas~\cite{widmann1995, radtke2000}. 
The present work also demonstrates the application of the Compton polarimetry technique, which provides advantages over Bragg polarimetry, previously used for DR polarization measurements at EBITs~\cite{shlyaptseva1997, shlyaptseva1998, shlyaptseva1999}, by measuring both the degree and the angle of polarization and by being applicable in a broad range of energies. 
This technique can also be applied to investigate polarization properties of other electron--ion collision processes such as resonant recombination~\cite{joerg2015,chen15} and excitation, radiative recombination~\cite{ma2003,tashenov2006, tashenov2014}, electron impact excitation~\cite{beiersdorfer1996, takacs1996, beiersdorfer1997, beiersdorfer1999, nakamura2001, robbins2004, robbins2006, gumberidze2013}, nuclear-field-induced excitation~\cite{gumberidze2011} and higher-order resonant recombination processes~\cite{beilmann2011, beilmann2013}. 
Due to its high sensitivity, Compton polarimetry can also be used for studies of weak radiative transitions, or, alternatively, for high accuracy studies of stronger transitions, in particular of those sensitive to effects of quantum electrodynamics~\cite{tong2015}.

%
%

%
%

\begin{acknowledgments}
This work was supported by the Deutsche Forschungsgemeinschaft (DFG) within the Emmy Noether program under Contract No. TA 740 1-1 and by the Bundesministerium f\"ur Bildung und Forschung (BMBF) under Contract No. 05K13VH2. 
\end{acknowledgments}


%

\end{document}